\documentclass{PoS}
\usepackage{bm}
\usepackage{amsmath}

\title{Model study of the sign problem in a mean-field approximation}

\ShortTitle{Model study of the sign problem in a mean-field
approximation}

\author{Yoshimasa Hidaka\\         RIKEN BNL Research Center, Brookhaven National
 Laboratory, Upton, New York 11973, USA\\}

\abstract{We study the sign problem of the fermion determinant at nonzero
baryon chemical potential. 
For this purpose we apply a simple model derived from Quantum
Chromodynamics, in the limit of
large chemical potential and mass. 
For SU(2) color, there is no sign problem and the mean-field
approximation is similar to data from the lattice.
For SU(3) color the sign problem is unavoidable, even in a mean-field
approximation. 
We apply a phase-reweighting method, combined with the mean-field
approximation, to estimate thermodynamic quantities. 
We also investigate the mean-field free energy using a saddle-point
approximation \cite{Fukushima:2006uv}.}

\FullConference{The XXV International Symposium on Lattice Field Theory\\
		 July 30-4 August 2007\\
		 Regensburg, Germany}

\begin{document}

\section{Introduction}
\label{sec:intro}
One of the current thrusts of hadronic physics is to understand 
the extreme conditions at high temperature and/or densities. 
The RHIC experiments reveal interesting
features of a Quark-Gluon Plasma (QGP) phase above the critical temperature. 
In the core of a dense neutron star,
various color superconductors should exist. Lattice
calculations, based on Monte-Carlo
simulation, are a powerful tool for the nonperturbative analysis of QCD.
At zero baryon density, results from the lattice
provide us with fundamental information, such as the phase
transition temperature
$T_\text{c}$ \cite{Tc}, the equation of state \cite{Aoki:2005vt},
susceptibilities \cite{Bernard:2004je}, the behavior of
correlation functions, and so on.

On the other hand, at nonzero quark density lattice simulations have a
serious sign problem: 
the quark determinant is complex in the presence of
the baryon chemical potential, so Boltzmann weights are complex,
and importantance sampling fails.
There are ways to overcome the problem, including reweighting
\cite{reweighting}, Taylor expansion in the chemical potential
\cite{Taylor}, and analytical continuation from imaginary values of the
chemical potential \cite{imaginary}. These methods are applicable
when the chemical potential is small, and the temperature high.  The sign
problem is even more intractable
at low temperature and high density. In this work we study the sign
problem in a toy model using a mean-field approximation.

Let us first see how the sign problem arises. The quark determinant is
\begin{equation}
 \det\mathcal{M}(\mu_q)\equiv\det [\gamma_\mu D^\mu +\gamma_4\mu_q+m_q],
\end{equation}
where $D^\mu\equiv\partial^\mu-igA^\mu$ is the covariant derivative, $m_q$
is the quark mass and $\mu_q$ is the quark chemical
potential. The quark determinant is complex except for
$\mu_q = 0$, due to a lack of gamma-five Hermiticity of 
$\mathcal{M}(\mu_q)$: $\det\mathcal{M}(\mu_q) = \det
\gamma_5\mathcal{M}(\mu_q)\gamma_5 = \{\det\mathcal{M}(-\mu_q)\}^*$. 
In and of itself, a complex quark determinant is not necessarily fatal.
While the quark 
determinant is complex for any given
$A_\mu$, the functional integral over $A_\mu$
is real for real observables. This is seen from the relation
\begin{equation}
\det 
 [\gamma_\mu(\partial^\mu-ig(A^\mu)^{C}+\gamma_4\mu_q+m_q)]=\{\det[
\gamma_\mu D^\mu+\gamma_4\mu_q+m_q]\}^*.
\end{equation}
The real part of the quark determinant is
$C$-even, and the imaginary part, $C$-odd. For a $C$-even ($C$-odd)
observable, then, the imaginary (real) part of the determinant vanishes
after integration over
$A_\mu$. Accordingly the real problem is that the contribution of the quark
determinant changes sign, depending upon $A_\mu$, and there is no known
method to replace importance sampling.

\section{Model}
We analyze a simple model to see the sign problem in the mean-field
approximation. The model
is obtained by taking double limit of heavy mass, $m\to\infty$, and large
chemical potential, $\mu\to\infty$,
keeping the ratio $\epsilon\equiv(e^{\mu_qa}/2m_qa)^{N_\tau}$
fixed. In the heavy
quark limit all excited quarks are static, while
antiquarks are suppressed at nonzero quark density, so that in the end,
the quark determinant can be rewritten in terms of the
Polyakov loop, 
\begin{equation}
e^{-S_\text{f}[L]}\equiv\det[\gamma_\mu 
 D^\mu+\gamma_4\mu_q+m_q]\to[\det(1+\epsilon L)]^{N_\text{f}/4},
\end{equation}
where $L(\vec{x}) = \prod_{\vec{x}_4}U_{\vec{x}_4}(\vec{x})$
is the Wilson line, which is a color matrix.  Here we 
set $N_\text{f} = 4$ in order to avoid the rooting problem with staggered
quarks. The determinant can be explicitly calculated:
\begin{equation}
\det{\mathcal{M}}=
\left\{
\begin{array}{lc}
\displaystyle
\prod_{\vec{x}}(1+\epsilon^2+2\epsilon\ell) & \text{for SU(2)}\\
\displaystyle
\prod_{\vec{x}}(1+\epsilon^3+3\epsilon\ell+3\epsilon^2\ell^*) & \text{for
SU(3)}
\end{array}
\right.,
\end{equation}
where the Polyakov loop is the trace of the Wilson
line, $\ell(\vec{x}) = \text{tr}L(\vec{x})/N_\text{c}$
in the fundamental representation. One can easily check that 
SU(2) color does not have the sign problem
since the Polyakov loop is always real, $-1\leq\ell\leq1$. For SU(3)
case the Polyakov loop $\ell$ is complex valued, and
the determinant is complex. We analyze these two cases in the
next section.

For gluons, we take a simple action with nearest
neighbor interactions between Polyakov loops,
\begin{equation}
S_\text{g}=-N_\text{c}^2J\sum_{\text{n.n.}}\ell(\vec{x})\ell^*(\vec{y}).
\end{equation}
$J$ is a parameter which can be interpreted as the temperature
of the system. In the strong
coupling expansion, $J$ is related to the true temperture, $T$, through $J =
\exp[-\sigma a/T]$, where $\sigma$ is the string tension.
It is known that this action reproduces the gross features of the phase
transition without quarks;
i.e., a second-order phase transition for SU(2) color, and a
first-order phase transition for 
SU(3) color.  In this
work we leave
$J$ as a free parameter.

\section{Mean-field approximation}
At nonzero temperature the free energy is related to the functional integral as
\begin{equation}
e^{-\beta Vf}=Z(\beta,\mu)=\int DL\exp(-S),
\end{equation}
$\beta=1/T$. Assuming that the action $S$ is real, 
$\exp(-S)$ is positive 
semidefinite, and in a mean-field approximation the free energy is:
\begin{eqnarray}
e^{-\beta Vf}&=&\int
DL\exp(-S_\text{mf}(x)+S_\text{mf}(x)-S)=Z_{\text{mf}}(\beta)\langle\exp(S_
\text{mf}(x)-S)\rangle_{\text{mf}}
\nonumber\\
&\geq& Z_\text{mf}(\beta)\exp(\langle 
S_\text{mf}(x)-S\rangle_\text{mf})\equiv e^{-\beta V f(x)}.
\label{eq:freeenergy}
\end{eqnarray}
The average $\langle\cdots\rangle_\text{mf}$ is taken with respect to
mean-field action $S_\text{mf}(x)$; $x$ is a parameter
of mean-field theory. In eq.~(\ref{eq:freeenergy}), the first line  
is an identity, while the second line follows from
Jensen's inequality,
$\langle(\exp\mathcal{O})\rangle \geq \exp(\langle\mathcal{O}\rangle)$. 
The mean-field free energy is larger or equal than the exact free energy 
for any $x$. The inequality ensures that the $f(x)$ has a minimum at
$x=x_0$, 
\begin{equation}
\left.\frac{\partial f(x)}{\partial 
 x}\right|_{x=x_0}=0,\hspace{0.5cm}\text{and}\hspace{0.5cm}\left.
\frac{\partial^2f(x)}{\partial 
 x^2}\right|_{x=x_0}\geq 0~.
\end{equation}
Our ansatz for the mean-field action is
\begin{equation}
S_\text{mf}[L]\equiv-\frac{x}{2}\sum_{\vec{x}}[\ell(\vec{x})+\ell^*(\vec{x}
)].
\end{equation}
Once $S_\text{mf}[L]$ is known, and $x$ determined, the expectation value
of any observable $\mathcal{O}[L]$ 
is given by integrating with repsect to the Wilson line, $L$, over
the group measure, with the action of mean-field theory,
$S_\text{mf}[L]$:
$\langle\mathcal{O}[L]\rangle\simeq\langle\mathcal{O}[L]\rangle_\text{mf}$.
If the action is not real, inequality is not ensured, and convexity is
violated. This how the sign problem manifests itself
in a mean-field approximation, and occurs for three or more colors.
Charge-conjugation symmetry is
violated at non-zero quark density, which is seen from
$\langle\ell\rangle\neq\langle\ell^*\rangle$. Here we note that
both $\langle\ell\rangle$ and $\langle\ell^*\rangle$ are real valued,
from the argument in
Sec.\ref{sec:intro}. This difference has been observed in both lattice
simulations \cite{Taylor} and in other models \cite{Dumitru:2005ng}. It
is necessary to extend the mean-field ansatz (3.4) to include two variables,
$x$ and $y$, in order to represent
the difference between $\langle\ell\rangle$ and
$\langle\ell^*\rangle$:
\begin{align}
S_\text{mf}=-\frac{x}{2}\sum_{\vec{x}}[\ell(\vec{x})+\ell^*(\vec{x})]-\frac
{y}{2}\sum_{\vec{x}}[\ell(\vec{x})-\ell^*(\vec{x})].
\end{align}
While the mean-field action is complex, $x$ and $y$ are real, so
that after integrating over $L$, the
free energy $f_\text{mf}(x,y)$ is a real function of $x$ and $y$.
Their values are then determined by requiring that the free energy
is a stationary point.  At $\mu_q \neq 0$, $y \neq 0$.
It turns out that about the stationary point,
while the free energy $f_\text{mf}(x,y)$ is
minimal in the $x$ direction, it is maximal
in the $y$ direction. That is, the solution is a
saddle-point in $x$ and $y$, 
consistent with Ref. \cite{Dumitru:2005ng}.

The phase reweighting method is one way to deal with the complexity of
the action.  The magnitude of the quark determinant is $C$-even,
while its phase is $C$-odd.
Accordingly, the quark action is
\begin{equation}
S_\text{f}=S_\text{f}^{\text{mag}} +i\Theta[L],
\end{equation}
where
\begin{eqnarray}
S_\text{f}^{\text{mag}}&=&-\sum_{\vec{x}}\ln|1+\epsilon^3+3\epsilon\ell+3
\epsilon\ell^*|,\\
\Theta[L]             
&=&-\sum_{\vec{x}}\arg(1+\epsilon^3+3\epsilon\ell+3\epsilon^2\ell^*).
\end{eqnarray}
With these definitions the expectation value of
$\mathcal{O}[L]$ is
\begin{equation}
\langle\mathcal{O}[L]\rangle\simeq\langle
\mathcal{O}[L]e^{-i\Theta[L]}\rangle_\text{mf}\Big/\langle
e^{-i\Theta[L]}\rangle_\text{mf}.
\end{equation}
Here $S_\text{mf}[L]$ or $x$ is fixed from the free energy with the action
$S_\text{g}[L]+S^\text{mag}_\text{f} [L]$, so that $x$ encompasses
the information of $S^\text{mag}_\text{f} [L]$ implicitly. This scheme is
the same as what has been adopted in the
lattice simulations of Ref. \cite{Blum:1995cb}. We compare these two
methods for SU(3) color, and find qualitatively similar behavior.

\section{Results}
\begin{figure}
\includegraphics[width=\linewidth]{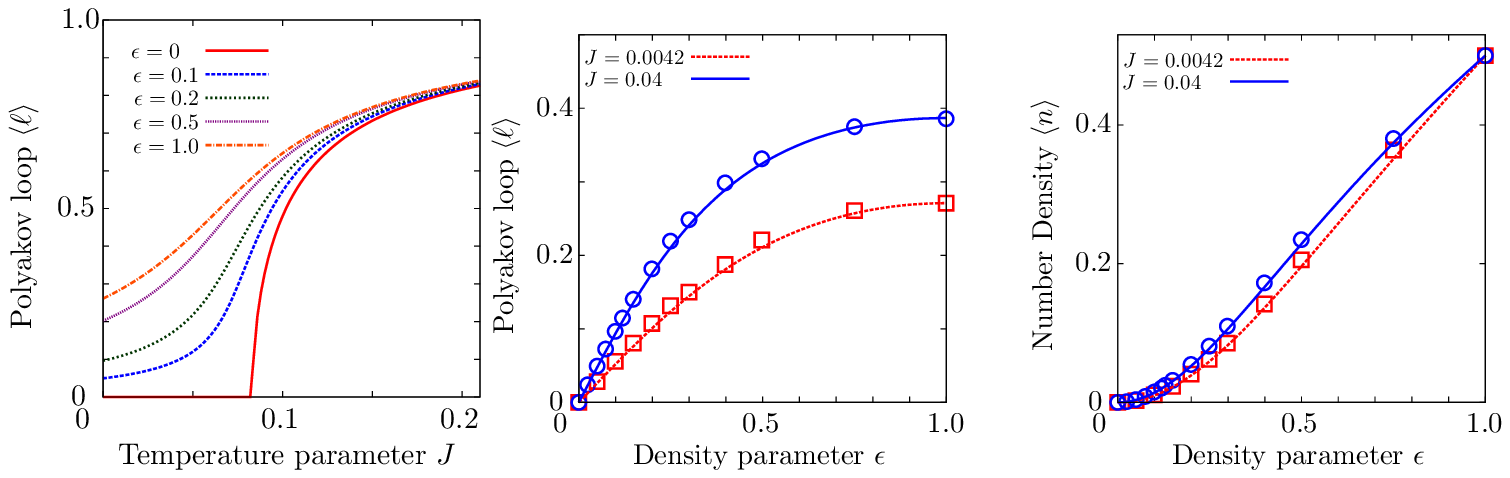}
\caption{For SU(2),
comparison of the model to lattice data,
Fig. 2 of Ref. \cite{Blum:1995cb}.  
Left: the Polyakov loop versus the temperature
parameter.  Center: the Polyakov loop versus the density parameter.
Right: the number density versus the density parameter.
}
\label{fig:1}
\end{figure}
\subsection{SU(2)}
We first consider SU(2) color, to see how a
mean-field approximation works when there is no sign problem.
We look for a phase transition by considering how the Polyakov loop
changes as $J$ increases.
In the pure glue theory, the deconfining phase transition is
known to be of second order for two colors, in
the universality class of the Ising model.
In our model, at $\epsilon = 0$ there is a continuous transition at 
$J = J_\text{c}\simeq0.083$, 
as indicated by the
solid curve in the left figure of Fig.~\ref{fig:1}.  The presence of
dynamical quarks acts on the Polyakov loop
variable as an external field which breaks the center symmetry. In fact, the
results at $\epsilon \neq 0$ in the left
figure of Fig.~\ref{fig:1} indicate not a true phase
transition, but only crossover.

Our mean-field outputs are to be compared with the lattice simulations in
Ref. \cite{Blum:1995cb}: the center
and left figures in Fig.~\ref{fig:1} correspond to Figs.~1 and 2
of Ref. \cite{Blum:1995cb}, respectively. We cannot
expect exact agreement, because our ansatz for the pure gluonic
action $S_\text{g}[L]$ is only a
crude approximation of QCD, and in any case,
we neglect the renormalization of the Polyakov loop in
a mean-field analysis. Nevertheless, the agreement turns out to be
surprisingly good, beyond naive
expectation, if the parameter $J$ is treated as an adjustable parameter
as a fitting parameter.
In this way, we
fix $J = 0.0042$ and $J = 0.04$ to reproduce the SU(2) Polyakov loop 
{\it only} at $\epsilon = 1$ for $4/g^2 = 2.0$
and $4/g^2 = 1.5$, respectively. We stress that we do not use the
data of the Polyakov
loop at $\epsilon\neq 1$, nor the results on the
number density.  Nevertheless, as clearly seen from 
Fig.~\ref{fig:1}, our numerical results fit all of the
lattice data remarkably well. We conclude
from this that the main corrections to our ansatz (2.3)
can be represented by a shift in the parameter $J$.
This gives us confidence in using a mean-field approximation for this
problem.

\subsection{SU(3)}
\begin{figure}
\includegraphics[width=\linewidth]{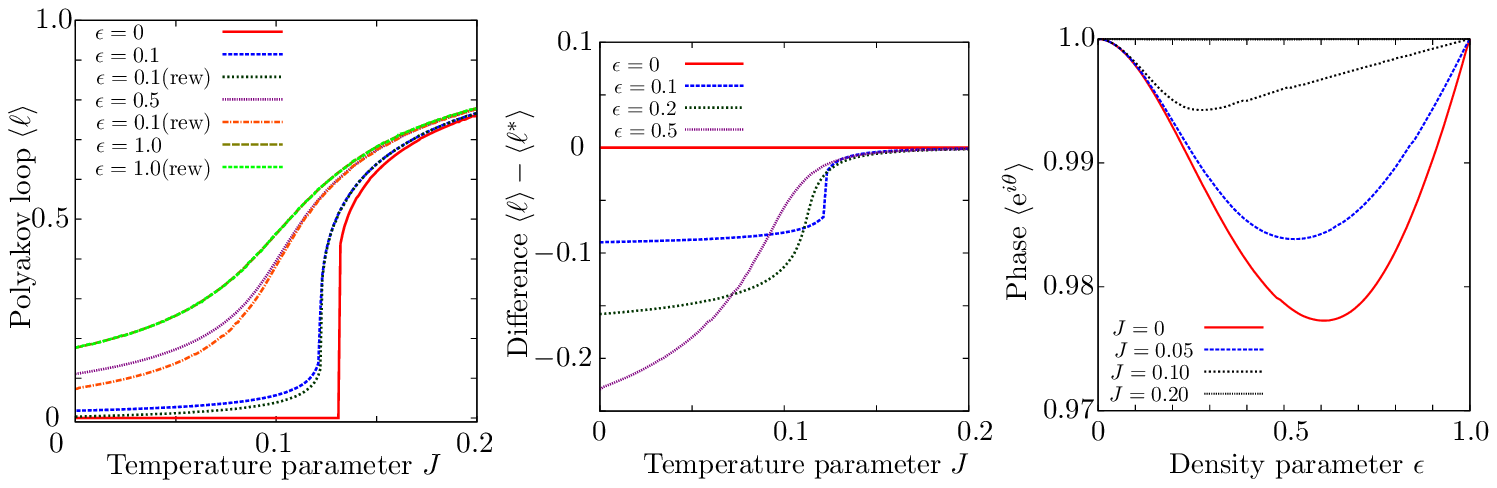}
\caption{
For SU(3), comparison of the model to lattice data,
Fig. 2 of Ref. \cite{Blum:1995cb}.  
Left: comparison of the Polyakov loop between phase reweighting
and the saddle
point approximation.  Center: difference between
$\langle\ell\rangle$ and $\langle\ell^*\rangle$, versus the
temperature parameter.  Right: expectation value of the phase of the quark
determinant versus the density parameter.}
\label{fig:2}
\end{figure}
We next consider SU(3) color.  The Polyakov loop is compared in
the phase 
reweighting method, and the saddle point approximation, in the left
figure in Fig.~\ref{fig:2}; this is 
to be compared with the lattice results from Fig. 7 of Ref.
\cite{Blum:1995cb}. We find a first-order phase
transition for $\epsilon = 0$ at $J = J_\text{c} = 0.132$ and for $\epsilon
= 0.1$ at $J = J_\text{c} = 0.123$.
Nonzero $\epsilon$ smears the transition, so that it 
eventually ceases to be of first-order.
The line of first-order transitions ends with a second order transition,
which is then a 
critical end-point.  For larger values of $\epsilon$ there is
only crossover. The global picture is consistent with 
results from a Potts model.  In
the left figure in Fig.~\ref{fig:2} one sees that both reweighting,
and the saddle-point approximation,
have qualitatively the same behavior for the expectation value of the Polyakov
loop.

At $\mu_q\neq0$, $\langle\ell\rangle \neq \langle\ell^*\rangle$.
The observable $\langle \ell-\ell^* \rangle$ is $C$-odd, where
the imaginary part of the fermion determinant is responsible for this
difference. In the
center figure of Fig.~\ref{fig:2} we present our numerical results for the
difference $\langle\ell\rangle-\langle\ell^*\rangle$ as a function of
$J$. The difference is trivially zero at $\epsilon = 0$ and $\epsilon = 1$
where the fermion determinant is real. As long
as the density parameter stays smaller than $\epsilon \sim0.5$, a larger
density parameter $\epsilon$ leads to a bigger
difference. For example, at $\epsilon = 0.5$ we find
$\langle\ell\rangle -\langle\ell^*\rangle= -0.076$, which is
comparable to $\langle\ell\rangle= 0.073$.

One can intuitively understand why $\langle\ell^*\rangle$ is greater than
$\langle\ell\rangle$ at nonzero $\mu_q$, as seen in \cite{Taylor}. It is
because at nonzero quark density, 
the presence of quarks enhances the screening of
antiquarks, so that an antiquark costs less energy \cite{Taylor}.

Finally we present the results for the expectation value of the phase
factor of the quark
determinant, $e^{-i\Theta}$. We plot $\langle e^{-i\theta}\rangle$ as a
function of $\epsilon$ in Fig.~\ref{fig:2}, where $\theta$ is the phase at
each lattice site; 
$\theta\equiv-\arg(1+\epsilon^3+3\epsilon\ell+3\epsilon^2\ell^*)$, i.e.
$\Theta = \sum_{\vec{x}}\theta$.
Comparing it with Fig. 9 in Ref. \cite{Blum:1995cb},
we see that our results qualitatively reproduce the lattice
data. For more quantitative agreement, 
we approximate the phase factor by
$\langle e^{-i\theta}\rangle^{216}$,
taking the lattice volume of $6^3 = 216$ from \cite{Blum:1995cb}. For instance,
our $J = 0$ result has a minimum at $\epsilon = 0.61$ where $\langle
e^{-i\theta}\rangle\simeq0.977$, 
while we obtain $0.977^{216} = 0.0066$.
The minimum value in Fig. 9 of Ref. \cite{Blum:1995cb} is 
$\approx 0.01$, which is close to our value.

\section{Summary}
We have explored a simple model applicable in the limit of heavy quark mass
and large chemical potential, and seen how
the sign problem, at nonzero quark density, manifests itself in a 
mean-field approximation.  All results from mean-field theory
are reasonable, and are in quantitative agreement with
lattice data.
There is no sign problem for SU(2), and we find that a mean-field
approximation works well, for both the quark number density and the
Polyakov loop.

For SU(3), we compared two methods, a saddle point approximation with a
complex action, and phase reweighting.
We find that both methods give qualitatively the same behavior for
the expectation value of the Polyakov loop. The complex action implies that
$\langle\ell\rangle \neq \langle\ell^*\rangle$, which is also seen with
phase reweighting. We computed
$\langle\ell\rangle-\langle\ell^*\rangle$, as a function of the 
parameters for temperature and density.

While it may appear odd to find a saddle point in the mean field approximation,
it is known that this happens in other theories,
such as for a nonlinear sigma model.  There, 
the constraint of the nonlinear model is
eliminated by introducing a new field.
The effective action, including the constraint field, is complex, so
that the stationary points thereof are true saddle points.  
For QCD, the $A_0$ field is a constraint field which imposes 
Gauss' law.  This may provide a clue to resolving the sign problem with
dense quarks.

\end{document}